\documentclass[twocolumn,showpacs,prb]{revtex4}
\usepackage[dvips]{graphicx}

\newcommand{\mvec}[1]{{\mathbf{#1}}}
\newcommand{\Vsub}{V_{\mathrm{MF}}}

\newcommand{\Esub}{{\mathcal E}^{\mathrm{MF}}}
\newcommand{\Ecoup}{{E^{\mathrm{coup}}}}
\newcommand{\Qmax}{Q_{\mathrm{max}}}
\newcommand{\usqr}{{\langle u^2\rangle}}
\renewcommand{\d}{{\mathrm{d}}}
\renewcommand{\log}{{\:\mathrm{ln}}}

\newcommand{\sdot}{{\!\,\cdot\,\!}}

\renewcommand{\rq}{{small-harmonics }}

\begin{document}

\title{Vortex matter in layered superconductors without
Josephson coupling:\\numerical simulations within a mean-field approach}

\author{Hans Fangohr}

\affiliation{
School of Engineering Sciences and Department of Physics and Astronomy, \\
University of Southampton, Southampton, SO17 1BJ, United Kingdom
}

\author{Alexei E. Koshelev}
\affiliation{
Materials Science Division, Argonne National Laboratory\\
 9700 South Cass Avenue, Argonne, Illinois 60439, United States.
}

\author{Matthew J. W. Dodgson}
\affiliation{
Theory of Condensed Matter Group, Cavendish Laboratory  \\
University of Cambridge, Madingley Road, Cambridge, CB3 0HE, United Kingdom
}

\begin{abstract}
  We study vortex matter in layered superconductors in the limit of
  zero Josephson coupling. The long range of the interaction between
  pancake vortices in the $c$-direction allows us to employ a
  mean-field method: all attractive inter-layer interactions are
  reduced to an effective substrate potential, which pancakes
  experience in addition to the same-layer pancake repulsion. We
  perform numerical simulations of this mean-field model using two
  independent numerical implementations with different simulation
  methods (Monte-Carlo sampling and Langevin molecular dynamics).  The
  substrate potential is updated self-consistently from the averaged
  pancake density. Depending on temperature, this potential converges
  to a periodic profile (crystal) or vanishes (liquid). We compute
  thermodynamic properties of the system, such as the melting line,
  the instability line of the crystal, and the entropy jump across the
  melting transition. The simulation results are in good agreement
  with approximate analytical calculations.
\end{abstract}


\maketitle


\section{Introduction}
\label{sec:introduction}

The vortex state in type-II superconductors is a complex physical
system. Within the layered high-temperature materials, such as
Bi$_2$Sr$_2$CaCu$_2$O$_8$ (BSCCO) or YBa$_2$Cu$_3$O$_7$ (YBCO), vortex
lines can be understood as wiggling stacks of pancake
vortices.\cite{Artemenko90,Feigelman90,Buzdin90,Clem91,Blatter94} The
thermodynamic properties of the vortex state are determined by the
interaction between pancake vortices. There are two mechanisms of
pancake interaction: (i) electromagnetic interaction and (ii)
Josephson coupling. The electromagnetic interaction is mediated by
supercurrents circulating around each pancake, whereas the Josephson
coupling results from the energy cost due to a phase shift between the
superconducting order parameters in the neighboring layers.

To understand the phase diagram of high-temperature superconductors
and in particular the melting line of the vortex
lattice,\cite{Zeldov95,Schilling96} we need to gain an insight into
the behavior of vortex matter under a variety of experimental
conditions. In moderately anisotropic materials, such as YBCO, the
short-range Josephson coupling is the dominant inter-layer
interaction, and the vortices are well described as elastic
strings.\cite{Ryu96a,Nordborg97,Wilkin97,Nordborg98,Otterlo98,Olson00a}
In very anisotropic materials on the other hand, such as BSCCO, the
Josephson coupling is weak, and the long-range electromagnetic
interaction between the pancakes should be taken into account. In this
paper we consider very anisotropic materials in the absence of
Josephson coupling, and neglect pinning. Even after keeping only the
electromagnetic coupling, the problem remains a challenging one, due
to the long range of the interactions: the energy of electromagnetic
interaction between two pancakes depends logarithmically on the
separation along the layers and decays exponentially with the number
of layers between the pancakes. More specifically, the interaction is
repulsive between pancakes in the same layer, and attractive between
pancakes in different layers, and the decay length of the exponential
dependence is the London penetration depth $\lambda$, which is
typically $100$ times larger than the layer spacing $s$.
Approximately, this system has been investigated within the
density-functional theory. \cite{Sengupta91,Menon96,Cornaglia00}

For a numerical investigation of the system, one can in principle
simulate directly a stack of two-dimensional (2D) pancake systems
taking into account all of the interlayer interactions. However, the
computational challenge is that the interlayer attraction between
pancakes extends over a range of $2\lambda/s \sim 100-150$ layers. In
addition, realistic simulation of the melting transition requires at
least several hundred point vortices per layer. So far, direct
numerical investigations have been performed only on small systems
using about 10 layers and of the order of 100 vortices.
\cite{Reefman91,Kolton00,Kolton00a,Olson00,Olson00e,Olson01f} This is
not sufficient to describe realistically the vortex state in BSCCO.
With today's computational resources, it is not feasible to perform
realistic direct three-dimensional (3D) simulations of this system
because the necessary computational effort grows quadratically with
the number of layers.

\begin{figure}

  \centerline{\includegraphics[width=0.8\columnwidth]{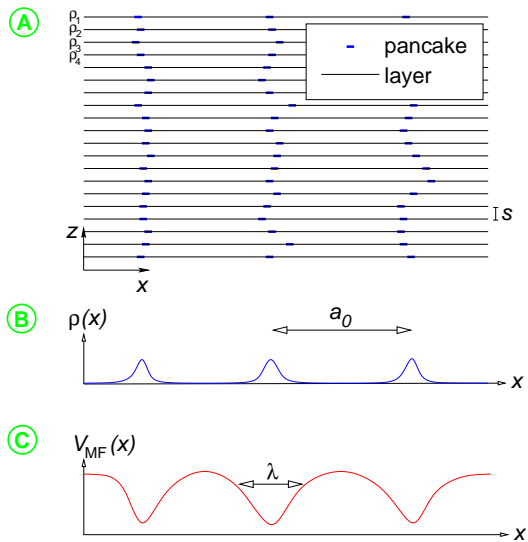}}
\caption[Substrate model]{
  Schematic representation of the substrate model.  \emph{(A)} The
  pancake positions $\rho_n(x)$ in each layer $n$, are \emph{(B)}
  averaged over the layers in order to obtain the averaged pancake
  density $\rho(x)$. From the average pancake density we compute
  \emph{(C)} the substrate potential $\Vsub(x)$, which is smeared over
  a length of the order of $\lambda$.  The vortex lattice spacing is
  $a_0$ and $s$ is the layer spacing.
  \label{fig:substrate-principle}
}

\end{figure}

Fortunately, one can benefit from the long range of the interlayer
coupling.  As the interlayer force on a pancake is the result of a sum
of a large number ($\sim 2\lambda/s$) of small contributions, it can
be calculated by a mean-field approach. The exact value of this force
is determined by the instantaneous pancake densities in the large
number of layers. In the crystal state the instantaneous density can
be decomposed into the average density, which is a periodic function
of the in-plane coordinates, and a fluctuating contribution. In the
mean-field approach to the interlayer interactions, one replaces the
instantaneous densities in the other layers by the average density.
This approach gives a quantitatively correct description of the
system, because due to the law of large numbers, the neglected force
from the fluctuating densities is typically smaller than the average
interlayer force by the factor $\sim \sqrt{s/\lambda}\ll 1$. The
calculation then takes the form of independent layers, with the
pancakes in each layer feeling an effective ``substrate
potential''.\cite{Dodgson00} This substrate potential is the
cumulative affect of the attraction of pancakes in all other layers as
illustrated in Fig.~\ref{fig:substrate-principle}. Pancakes within one
layer interact directly with each other, whereas the interaction with
pancakes in other layers is mediated via the substrate potential.
Thus, each layer is treated individually, until a new substrate
potential can be computed. This process is iterated, until the
substrate has converged to a steady solution. In this paper, we
present the first numerical implementations of this substrate model
and show results which we compare with the semi-analytic
approximations given in Ref. [\onlinecite{Dodgson00}].

\begin{figure}
  \begin{center}

    \raisebox{0.95cm}{\includegraphics[clip,width=3.5cm]{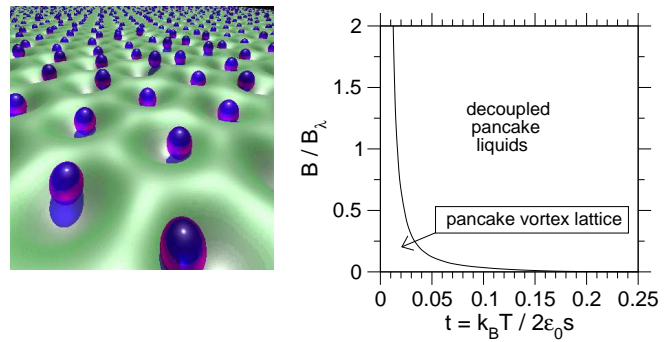}
      } \hfill
    \includegraphics[clip,width=4.8cm]{figure2b.eps}

    \caption{
      Left: Snapshot of the pancakes (visualized by spheres) placed
      onto the substrate potential (visualized as a surface) at
      $B=B_\lambda$ and $t=1/59\approx0.017$ just below the melting
      transition to demonstrate how the substrate potential constrains
      the pancake motion. Right: The phase diagram we have computed
      using the substrate method.
      \label{fig:pancakes-in-substrate}}
\end{center}
\end{figure}
We summarize this work in Fig.~\ref{fig:pancakes-in-substrate}. On the
left the central idea is visualized: pancakes experience attractive
inter-layer interactions through the substrate potential which
stabilizes the pancake crystal. On the right we show the computed
melting line separating a 3D pancake vortex lattice from decoupled 2D
liquids. We express magnetic induction in units of $B_\lambda\equiv
\Phi_0 / \lambda^2$, where $\Phi_0$ is the magnetic flux quantum, such
that the pancake-spacing in a triangular lattice is
$a_0=\sqrt{2/\sqrt{3}}\ \lambda \approx 1.07 \lambda $ at
$B=B_\lambda$.  We use a dimensionless temperature, $t$, which is the
ratio of the thermal energy $k_\mathrm{B}T$ to the prefactor $2 s
\epsilon_0$ of the logarithmic pancake-pancake interaction,
\begin{equation}
t  \equiv \frac{1}{\Gamma} \equiv
\frac{k_\mathrm{B}T}{2\epsilon_0s}, \label{eq:def-Gamma}
\end{equation}
where $\epsilon_0 = \Phi_0^2/(4\pi\mu_0\lambda^2)$, $\mu_0$ is the
vacuum permeability, and $s$ is the layer spacing. This allows to
compare our results with outcomes from 2D one-component Coulomb plasma
simulations,\cite{deLeeuw82,Caillol82,Choquard83} where frequently
$\Gamma = 1/t$ is used to express temperatures. At low fields, the
electro-magnetic attraction of range $\lambda \gg s$ between pancakes
in different layers stabilizes the 3D pancake-vortex lattice.
Increasing the magnetic field decreases the relative strength of the
inter-layer coupling. At high fields, $B\gg B_\lambda$, the long-range
repulsive interaction within the layer dominates, and the 3D pancake
lattice melts at a temperature close to the 2D melting temperature.

In Sec.\ \ref{sec:substrate model} we describe the substrate model in
detail, including three different methods for the efficient
computation of the substrate potential (Sec.~\ref{sec:numerical
  implementation}). The results, including the equilibrium phase
diagram, are shown in Sec. \ref{sec:results} before we conclude in
Sec. \ref{sec:conclusions}. The appendix gives a derivation of the
correlation correction to the free energy, and shows that our
mean-field approach should be accurate to order $s/\lambda$.

\section{Mean-field approach (Substrate model)}
\label{sec:substrate model}

\subsection{The mean-field inter-layer coupling}

The in-layer energy $ {\mathcal{E}}^{\text{in}}$ and the inter-layer
energy $ {\mathcal{E}}^{\text{inter}}$ of a system of
electromagnetically interacting pancakes in a layered superconductor
is, respectively,
\begin{equation}
  {\mathcal{E}}^{\text{in}} = \sum\limits_n E^\mathrm{in}_n
  = \sum\limits_{n} \frac{1}{2} \sum\limits_{j'\neq j} U (\mvec{R}_j^n - \mvec{R}_{j'}^n , 0)
\label{eq:in-layer-interaction}
\end{equation}
and
\begin{equation}
{\mathcal{E}}^{\text{inter}}\!
   = \frac{1}{2}\sum\limits_{n'\neq n}\sum\limits_{j, j'} U ( \mvec{R}_j^n - \mvec{R}_{j'}^{n'} , n-{n'} ).
\end{equation}
Indices $n$ and ${n'}$ count over layers, and $j$ and $j'$ over
pancakes in the layers, $\mvec{R}_j^n$ is the (2D) position of pancake
$j$ in layer $n$, and $U( \mvec{R},n)$ is the coupling energy for two
pancakes separated by a vector $( \mvec{R}, z )$, where $z=ns$, with
$s$ being the layer spacing. The $z$-axis is chosen perpendicular to
the layers.

The in-layer pancake interaction\cite{Clem91} is
\begin{eqnarray}
\label{eq:U_pan_inlayer}
U(\mvec{r},0) &= 2\epsilon_{0}s &\left( \left(1-\frac{s}{2\lambda}\right)\log\left(\frac{L}{r}\right)\right. \\
              &       & \phantom{\left)\right.}\left.+\frac{s}{2\lambda}\int_{r}^{\infty}\d r' \frac{\exp(-r^{\prime}/\lambda)}{r^{\prime}}\right)\nonumber
\end{eqnarray}
and the inter-layer interaction ($n\neq0$) is
\begin{eqnarray}
\label{eq:U_pan_interlayer}
U(\mvec{r},n) &\!=&-\frac{\epsilon_{0}s^{2}}{\lambda}\!\left( \exp\left( -\frac{ns}{\lambda}\right) \log\left(\frac{L}{r}\right)\right. \\
&&\phantom{x}\left.-\int_{r}^{\infty}\!\!\!\!\!\! \d r'\, \frac{\exp ( - \sqrt{r^{\prime2}+(ns)^{2}}/{\lambda})}{r^{\prime}}\right). \nonumber
\end{eqnarray}

Using
\begin{equation}
  \label{eq:delta-density}
\rho_{n}(\mvec{r}) = \sum_{j}\delta(  \mvec{r}-\mvec{R}^n_{j})
\end{equation}
 we rewrite
\begin{equation}
{\mathcal{E}}^{\text{inter}}= \frac{1}{2}\sum_{n\neq {n'}}\int \!\text{d}^2 r \:\text{d}^2 r^{\prime} \:
 \rho_{n} ( \mvec{r} ) \rho_{{n'}} ( \mvec{r}^{\prime}) U ( \mvec{r}-\mvec{r}^{\prime},n-{n'}).
\label{eq:Einter-step2}
\end{equation}
We separate pancake density fluctuations from the layer-average density
\begin{eqnarray}
\rho( \mvec{r})  &\equiv& \langle \rho_{n}(\mvec{r})  \rangle ,  \label{eq:def-average-rho} \\
\rho_{n}(\mvec{r}) &=& \rho(\mvec{r})  +\delta \rho_{n}(  \mvec{r} ), \label{eq:def-rho_n}
\end{eqnarray}
and obtain from (\ref{eq:Einter-step2})
\begin{eqnarray}
{\mathcal{E}}^{\text{inter}}&=&\frac{1}{2}\sum_{n\neq n^{\prime}}\int \! \d^2r\d^2r^{\prime}U(  \mvec{r}-\mvec{r}^{\prime},n-n^{\prime}) \label{eq:all-fluctuations}\\
&&\times\big[  \rho(  \mvec{r})  \rho( \mvec{r}^{\prime})
   +2\rho (  \mvec{r}^{\prime})  \delta\rho_{n}(  \mvec{r} )  +\delta\rho_{n} ( \mvec{r})  \delta\rho_{n^{\prime} }(  \mvec{r}^{\prime})  \big]. \nonumber
\end{eqnarray}
Because the difference $n-n'$ in the last sum extends over a very
large number of layers ($\sim \lambda/s$), a typical value of the
sum $\sum_{n^{\prime}}U(
\mvec{r}-\mvec{r}^{\prime},n-n^{\prime})2\rho (
\mvec{r}^{\prime})$ is larger than a typical value of the sum
$\sum_{n^{\prime}}U(\mvec{r}-\mvec{r}^{\prime},n-n^{\prime})\delta\rho_{n^{\prime}
}$ by the factor $\sim \sqrt{\lambda/s}$.
Again, the law of large numbers
allows us to neglect the last term in
(\ref{eq:all-fluctuations}) leading to the mean-field description
of the interlayer interactions.
A more precise justification is given in the Appendix, where the free energy
correction due to the correlation term is shown to be smaller than the
mean-field free energy by the factor $s/\lambda$.

Separating the pancake density into the average value $\rho = \langle
\rho_n(\mathbf{r})\rangle$ and a modulating part, we can split the
total magnetic coupling energy into two parts, each with a quite
different meaning. The part containing the average density does not
depend on temperature and formally diverges due to the logarithmic
term in (\ref{eq:U_pan_interlayer}). This divergence exactly
compensates a similar divergence in the in-plane energy. Within the
mean-field approach the part of the coupling energy sensitive to
density variations is finite only in the crystal state. In the liquid
state it vanishes.

For the mean-field interlayer energy $\Esub$ we obtain from
(\ref{eq:all-fluctuations})
\begin{eqnarray}
&\Esub\!\!& = \sum\limits_n E^\mathrm{MF}_n \\
& = & \frac{1}{2}\sum_{n\neq n^{\prime}}\int \!
\d^2r\d^2r^{\prime}U(  \mathbf{r}-\mathbf{r}^{\prime},n-n^{\prime}) \nonumber \\
&&\qquad \qquad \qquad \times  \left[  \rho(  \mathbf{r})  \rho(\mathbf{r}^{\prime})  +2\rho(\mathbf{r}^{\prime})
\delta\rho_{n}(\mathbf{r}) \right]  \nonumber \\
& = &\frac{1}{2}\sum_{n}\int \!
\d^2r\: \Vsub(\mathbf{r})\rho(\mathbf{r})
  +\sum_{n}\int \!  \d^2r\:\Vsub(\mathbf{r})\delta\rho_{n}(\mathbf{r})  \nonumber  \\
\label{eq:E-MF-version2}\\
& \stackrel{(\ref{eq:def-rho_n})}{=} &-\frac{1}{2}\sum_{n}\int \! \d^2r\:\Vsub(\mathbf{r}) \rho(\mathbf{r})
   +\sum_{n}\int \!  \d^2r\:\Vsub(\mathbf{r})\rho_{n}(  \mathbf{r}). \nonumber \\
\label{eq:E-MF}
\end{eqnarray}
The last term describes fluctuations in the fixed substrate potential
$\Vsub$,
\begin{eqnarray}
\Vsub(\mathbf{r})  & = & \int \!  \d^2r^{\prime}
\left[ \sum_{n\neq 0}U(  \mathbf{r}-\mathbf{r}^{\prime},n) \right]  \rho(
\mathbf{r}^{\prime})  \\
& = &\int \!  \d^2r^{\prime}\mathcal{U}(  \mathbf{r}-\mathbf{r}
^{\prime})  \rho(  \mathbf{r}^{\prime})
\label{eq:subs-potential-explicit-convolution}
\\
& \equiv &  ( {\mathcal U} * \rho ) (\mvec{r})
\label{eq:subs-potential}
\end{eqnarray}
with
\begin{equation}
\mathcal{U}(  \mathbf{r})  \equiv\sum_{n\neq0}U(\mathbf{r},n).
\label{eq:sum-over-layers}
\end{equation}
$\mathcal{U}(\mvec{r})$ is the interaction potential of a pancake
separated by $\mvec{r}$ from a stack of pancakes minus the
interaction of the (missing) pancake in the same layer
and is given by\cite{Clem91}
\begin{equation}
  \mathcal{U}(  \mathbf{r}) = 2 \epsilon_0 s K_0\left( \frac{r}{\lambda} \right) - U( \mvec{r}, 0 )
\label{eq:the-kernel}
\end{equation}
with $K_0(x)$ being a modified Bessel function of the second kind.
Ignoring terms of the order of $s/2\lambda$, the pancake-pancake
repulsion (\ref{eq:U_pan_inlayer}) simplifies to
\begin{equation}
  U(\mvec{r},0) = 2\epsilon_0s\log\left(\frac{L}{r}\right).
\end{equation}
In our calculations we find it useful to use the form in Fourier
space,\cite{FourierDef}
\begin{eqnarray}
  {\mathcal U }(\mvec{q}) &=& 4 \pi \epsilon_0 s \left( \frac{1}{\lambda^{-2}+q^2} - \frac{1}{q^2} \right) \\
&=& - 4 \pi \epsilon_0 s \frac{\lambda^{-2}}{q^2(\lambda^{-2}+q^2)}.
\label{eq:substrate-kernel-in-k-space}
\end{eqnarray}

\subsection{Algorithm}\label{sec:thesubstratemodel}

In principle, the substrate model can be implemented as follows:
\begin{enumerate}
\item Assume initial pancake densities $\rho_n(\mvec{r})$, for example a
  hexagonal lattice in each layer $n$.
\item \label{item:algo-beginning} Average the pancake density
  $\rho_n(\mvec{r})$ over all layers to obtain $\rho(\mvec{r})$,
  (\ref{eq:def-average-rho}).

\item Compute the substrate potential $V_\mathrm{MF}(\mvec{r})$,
  (\ref{eq:subs-potential}), by
  convoluting
  the substrate interaction
  kernel $\mathcal{U}(\mvec{r})$, (\ref{eq:the-kernel}), with the
  average pancake density $\rho(\mvec{r})$
\begin{equation}
  \label{eq:subs-potential-convolution}
  \Vsub(\mathbf{r}) = ( {\mathcal U} * \rho ) (\mvec{r}).
\end{equation}
\item \label{item:algo-MC-LD-simulation} For each layer $n$ compute
  the pancake distribution $\rho_n(\mvec{r})$ using Monte Carlo or
  Langevin dynamics simulations. The total energy for layer $n$
  contains the direct pancake-pancake interaction within the layer
  (\ref{eq:in-layer-interaction})
  \begin{equation}
    E^\mathrm{in}_n = \frac{1}{2}\sum\limits_{j'\neq j} U (\mvec{R}_j^n - \mvec{R}_{j'}^n , 0),     \label{eq:in-layer-energy-for-one-layer}
  \end{equation}
  and the relevant interaction with pancakes in other layers via the substrate potential (\ref{eq:E-MF})
  \begin{eqnarray}
    \qquad E^\mathrm{MF}_n \! &\stackrel{(\ref{eq:E-MF})\!}{=}& -\underbrace{\frac{1}{2}\int \! \d^2r\:\Vsub(\mathbf{r}) \rho(\mathbf{r})}_\Ecoup +\int \! \d^2r\:\Vsub(\mathbf{r})\rho_{n}(\mathbf{r}) \nonumber \\
&\stackrel{(\ref{eq:delta-density})\!}{=}& -\Ecoup + \sum\limits_{j} V_{\mathrm{MF}}(\mathbf{R}_j^n),
  \end{eqnarray}
  
  $\Ecoup$ is constant for a given $\rho(\mvec{r})$ and can therefore
  be ignored within the Monte Carlo/Langevin simulation as it only
  shifts the energy scale.

\item Go to \ref{item:algo-beginning}, until $\Vsub$ (or $\rho$) has
  converged.
\end{enumerate}
Since the substrate potential $\Vsub$ in step
\ref{item:algo-MC-LD-simulation} is the same for all layers, we can
compute $\rho_n(\mvec{r})$ for many Langevin-dynamics time-steps (or
Monte-Carlo sweeps) rather than many layers.  Therefore, in order to
obtain the averaged pancake density $\rho(\mvec{r})$ in step
\ref{item:algo-beginning}, we average over time-steps (or sweeps)
computed in one layer rather than averaging over layers.

Using the substrate potential, we reduce the solution of the 3D
problem to performing one 2D simulation in the presence of the
iteratively refined substrate potential.

\subsection{Numerical implementation}
\label{sec:numerical implementation}

We exploit the convolution theorem and compute the substrate
potential in Fourier-space\cite{FourierDef}
\begin{eqnarray}
V_{\mathrm{MF}}(\mathbf{r}) &\stackrel{(\ref{eq:subs-potential})}{=}&  ( {\mathcal U} * \rho ) (\mvec{r})\\
&=& \int\!\! \frac{\d^2 q}{(2\pi)^2} \:\:{\mathcal U}(\mvec{q}) \rho(\mvec{q}) \exp(i\mvec{q}\sdot\mvec{r})
\label{eq:implementation-Vsub-convolution}
\end{eqnarray}
using the analytical Fourier transform ${\mathcal U}(\mvec{q})$ as
given in (\ref{eq:substrate-kernel-in-k-space}), and the numerically
computed
\begin{equation}
  \rho(\mvec{q}) = \int \!\!\d^2r  \:\:\rho(\mvec{r}) \exp(-i \mvec{q}\sdot\mvec{r} ).
\label{eq:implementation-rho-q}
\end{equation}
This has two advantages: firstly, we do not cut off the interaction
kernel ${\mathcal U}$ within the simulation cell as would be the case
in the real-space convolution.  Secondly, this is numerically more
efficient than performing the convolution
(\ref{eq:subs-potential-explicit-convolution}) directly.

We have used three different methods for computing
$V_{\mathrm{MF}}(\mathbf{r})$ numerically.

\subsubsection{The full method}
\label{sec:full-method}

The ``full method'' computes the substrate potential $V_\mathrm{MF}$
using the full spectrum $\rho(\mvec{q})$ of Fourier-components of the
average pancake density $\rho(\mvec{r})$ as shown in
(\ref{eq:implementation-Vsub-convolution}). In our simulations we use
a resolution of $\approx 100^2$ grid-cells per pancake in order to
compute $\rho(\mvec{r})$ as an average over time-steps/sweeps. This
results in reciprocal lattice vectors up to magnitudes of $\approx 100
Q_0$, where $Q_0=4\pi/(\sqrt{3}a_0)$, because
$|\mvec{Q}_\mathrm{max}|/Q_0 \approx 2\pi/(\Delta x Q_0) \approx a0/
\Delta x \approx 100$. The necessary discrete Fourier-transform of
$\rho(\mvec{r})$, and the inverse transform of
$V_\mathrm{MF}(\mvec{q})={\mathcal U}(\mvec{q}) \rho(\mvec{q})$ can be
done efficiently using an implementation of the Fast Fourier
Transform.\cite{Frigo98}

We pre-compute the substrate potential $V_\mathrm{MF}(\mvec{r})$ on a
mesh and interpolate subsequently for intermediate pancake positions
while performing Langevin dynamics in the fixed substrate. We compute
a new substrate every 200,000 time-steps. It is important to average
over so many time-steps to reduce density fluctuations (due to poor
statistics) in the pancake histogram, which would result in a deformed
substrate potential.

Note that $\rho(\mvec{r})$ and $\rho(\mvec{q})$ are discretized out of
numerical necessity to compute a histogram but not for conceptual
reasons.

\subsubsection{The Fourier-filtered method}\label{sec:fourier-filtered-method}

The average density $\rho(\mvec{r})$ should be a periodic function,
which can be represented by a discrete set of Fourier components.
Therefore, the second method uses only a subset $\mvec{Q}_\mu$ of the
Fourier components $\mvec{q}$ to represent $\rho(\mvec{q})$
\begin{equation}
  \rho^\mathrm{FF}(\mvec{q}) = (2\pi)^2\sum\limits_\mu \rho_{\mvec{Q}_\mu} \delta^2(\mvec{q}-\mvec{Q}_\mu)
\label{eq:sparse-method-rho_Q}
\end{equation}
which we determine from the maxima of the structure factor and
\begin{equation}
  \rho_{\mvec{\mvec{Q}}} = \left\langle \frac{1}{L_xL_y}\sum \limits_j \exp(-i\mvec{R}_j^c \sdot \mvec{Q} ) \right\rangle_c
\label{eqn:rhoQ-average-over-configurations}
\end{equation}
with $L_xL_y$ being the area of the simulation cell. We average over
a set of configurations $c$ of pancake positions $\mvec{R}_j^c$
(either sweeps or time-steps) to compute $\rho_{\mvec{\mvec{Q}}}$.

Using $\rho^\mathrm{FF}(\mvec{r})=(2\pi)^{-2}\int\d^2q \:\rho^\mathrm{FF}(\mvec{q}) \exp(i
\mvec{q}\sdot\mvec{r})$ to present $\rho(\mvec{r})$, we Fourier-filter
$\rho(\mvec{r})$, and keep only the relevant components for the
computation of the periodic substrate.
We can write
\begin{eqnarray}
V^\mathrm{FF}_{\mathrm{MF}}(\mathbf{r}) &\stackrel{(\ref{eq:implementation-Vsub-convolution})}{=}&
\int\!\! \frac{\d^2q}{(2\pi)^2} \:\:{\mathcal U}(\mvec{q}) \rho^\mathrm{FF}(\mvec{q}) \exp(i\mvec{q}\sdot\mvec{r}) \\
 &\stackrel{~(\ref{eq:sparse-method-rho_Q})}{=}& \sum\limits_\mu  {\mathcal U}(\mvec{Q}_\mu) \rho_{\mvec{Q}_\mu} \exp(i\mvec{Q}_\mu \sdot\mvec{r})
\label{eq:Vsub_FF_1}\\
 &\stackrel{(\ref{eq:substrate-kernel-in-k-space})}{=}&
 - 4 \pi \epsilon_0 s \sum\limits_\mu  \frac{\rho_{\mvec{Q}_\mu}
\exp(i\mvec{Q}_\mu
\sdot\mvec{r})}{{Q}_\mu^2(1+\lambda^{2}{Q}_\mu^2)}.
\label{eq:sparse-method-final-equation-for-substrate}
\end{eqnarray}
This is equivalent to using the full-method, but setting
$\rho(\mvec{Q})=0$ if $\mvec{Q} \not \in \{\mvec{Q}_\mu\}$.

The advantage of the Fourier-filtered method is that we need to
average over less iterations before we can compute a new pancake
density, and subsequently a new substrate, because the substrate is
per construction periodic. Using the Fourier-filtered method we use
500 time-steps/sweeps for each substrate iteration.

It turns out that it is not necessary to take the average
(\ref{eqn:rhoQ-average-over-configurations}) over different
configurations but it is sufficient to use just one configuration
(\emph{i.e.} one time step or sweep):
\begin{equation}
  \rho_{\mvec{\mvec{Q}}} = \frac{ 1 }{L_xL_y}\sum \limits_j \exp(-i\mvec{R}_j\sdot \mvec{Q} ).
\label{eq:rho_Q_simple}
\end{equation}
Nevertheless, we run a simulation for 500 time-steps/sweeps with the
same fixed substrate potential to reduce re-computation of
$\rho_\mvec{Q}$, and to give the pancakes some time to explore the
system with a new substrate potential.

\subsubsection{The \rq (Fourier-filtered) approximation}

\begin{figure}
  \includegraphics[width=0.6\columnwidth,clip]{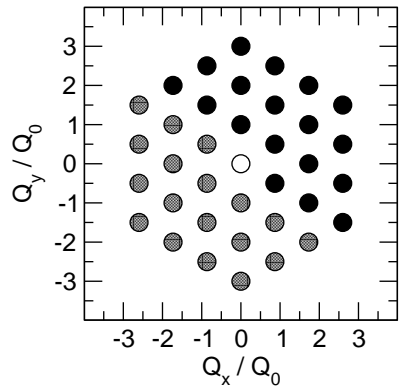}
\caption{
  Set of $\mvec{Q}_\mu$-vectors up to third order (\textit{i.e.} three
  ``shells'' around the origin) in reciprocal space used in the \rq
  Fourier-filtered method to compute $\rho_{\mvec{Q}_\mu}$. Due to the
  reality of $\rho(\mvec{r})$ we have $\rho(\mvec{q})=
  \bar{\rho}(-\mvec{q})$ and it is therefore sufficient to compute
  only half of the 36 coefficients $\rho_{\mvec{Q}_\mu}$.
  \label{fig:sparse-Qs}}

\end{figure}

In addition to Fourier-filtering $\rho(\mvec{r})$ we can speed up the
computation further because close to the melting temperature,
$\rho_{\mvec{Q}_\mu}$ decays quickly for higher-order $\mvec{Q}_\mu$
due to the Debye-Waller factor. We can estimate the reduction of
$\rho_{\mvec{Q}}$ due to the Debye-Waller factor
\begin{eqnarray}
\exp\left(-\frac{\usqr Q^2}{4}\right) =
\exp \left(-\frac{1}{4}\frac{\usqr}{a_0^2}\frac{16\pi^2 Q^2}{Q_0^2}\right)
\label{eq:Debye-Waller-factor}
\end{eqnarray}
where $Q_0 = 4\pi/(\sqrt{3}a_0)$ and $\usqr$ is the mean-squared
fluctuation displacement.  Depending on $\usqr$ we can ignore all
$\rho(\mvec{Q}_\mu)$ with $|\mvec{Q}_\mu| > Q'$. For all but the
smallest fields, we find close to the melting transition $\usqr/a_0^2
\approx 0.02 - 0.03$ (see Section~\ref{sec:latentheat}), and it is
sufficient to include up to $3^\mathrm{rd}$-order vectors
$\mvec{Q}_\mu$ in the summation in
(\ref{eq:sparse-method-final-equation-for-substrate}) as shown in
Fig.~\ref{fig:sparse-Qs}.

For the \rq Fourier-filtered method it is more efficient to evaluate
(\ref{eq:sparse-method-final-equation-for-substrate}) for each pancake
position occurring in the Langevin/Monte-Carlo simulation rather than
pre-computing $\Vsub$ on a mesh.

We demonstrate the equivalence of the full and the Fourier-filtered
method for the determination of the instability line in
Sec.~\ref{sec:convergence}, and we compare with the \rq
Fourier-filtered method in Sec.~\ref{sec:alpha}.

\subsection{Monte Carlo and Langevin dynamics simulations}

We have two independent implementations of the \rq Fourier-filtered
method: AEK has written a Monte Carlo simulation that is based on
energy evaluations, and HF has implemented a Langevin dynamics
simulation based on force calculations. The results of both
implementations agree perfectly.

We follow standard vortex-state simulation
techniques,\cite{Koshelev94,Fangohr01b} including periodic boundary
conditions for the in-plane interactions.  We use a smooth cut-off for
the vortex in-plane interactions. \cite{Koshelev94,Fangohr00} For the
Langevin dynamics simulations, we compute the substrate forces
numerically from the pre-computed mesh (Sec. \ref{sec:full-method})
for the full method and the Fourier-filtered method. For the \rq
Fourier-filtered method we use the analytical derivative of
(\ref{eq:sparse-method-final-equation-for-substrate}).  The Monte
Carlo simulations were only implemented with the \rq method.  If not
stated otherwise we use a system with 1020 pancakes.

\section{Results}
\label{sec:results}

\subsection{Time convergence of the substrate potential}
\label{sec:convergence}

\begin{figure}
  \begin{center}
     \includegraphics[width=1.0\columnwidth,clip]{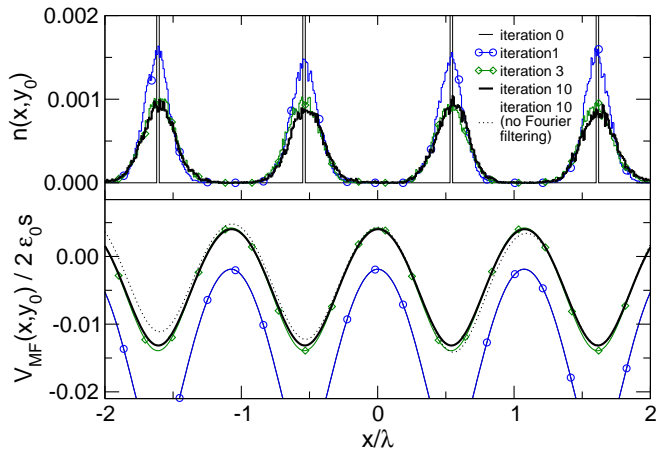}
    \caption{
      Convergence to pancake lattice at $B=B_\lambda$ and
      $t=1/59\approx0.017$.  \emph{Top:} pancake histogram $n(x,y_0)$
      taken along $y=y_0$.  \emph{Bottom:} substrate potential
      $\Vsub(x,y_0)$.
      \label{fig:convergence-to-solid}}

  \end{center}
\end{figure}

As described in Sec.~\ref{sec:thesubstratemodel}, we start each run
with a hexagonal pancake distribution corresponding to zero
temperature.  Fig.~\ref{fig:convergence-to-solid} shows results for
the Fourier-filtered method at $B=B_\lambda$ and at a temperature
$t=1/59\approx0.017$. The top plot shows a one-dimensional slice of
the 2D pancake histogram $n(x,y_0)$ taken along $x$ at $y=y_0$. The
histogram relates to the pancake density via $n(x,y) = \rho(x,y)\Delta
x \Delta y$ where $\Delta x$ and $\Delta y$ are the spacings of the
grid used to create the histogram.  For the $0^\mathrm{th}$
substrate-iteration we set the histogram to have narrow and high peaks
at the pancake equilibrium positions corresponding to delta-peaks in a
zero-temperature pancake density $\rho(\mvec{r})$. Based on this
initial pancake distribution, we compute the substrate potential,
$\Vsub(\mvec{r})$, for the first substrate-iteration, of which a
one-dimensional slice at $y=y_0$ is shown in the lower part of
Fig.~\ref{fig:convergence-to-solid}. Using this substrate potential,
we run the Langevin dynamics simulation for 200,000 steps which
results in the histogram for iteration 1 as shown in the upper plot of
Fig. \ref{fig:convergence-to-solid}. Based on these data, we compute
the substrate potential for iteration 2.  We iterate the substrate
re-computation until the substrate potential has reached a steady
state (after typically than 10 substrate-iterations).  The figure
demonstrates that the system converges quickly to a pancake solid at
this temperature below melting.

The dotted line in the lower part of
Fig.~\ref{fig:convergence-to-solid} shows a comparison substrate
potential for iteration 10 computed using the full method. While the
amplitude and width of the wells (and thus the resulting force) are
virtually identical to the Fourier-filtered data, the magnitude of the
substrate from the full method varies slightly. This is due to (long
wavelength) density fluctuations in the histogram data and reduces
further if one uses more time-steps for each substrate iteration.

\begin{figure}
  \begin{center}
        \includegraphics[clip,width=1.0\columnwidth]{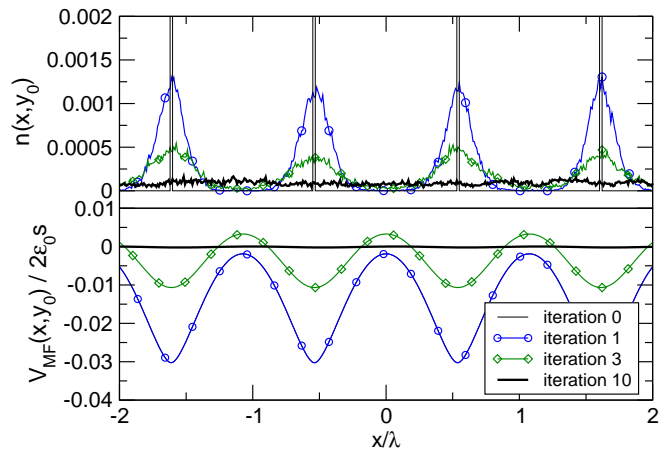}
    \caption{
      Convergence to pancake liquid. As
      Fig.~\ref{fig:convergence-to-solid} but at $t=1/50=0.02$ above
      the melting temperature at $B=B_\lambda$.
      \label{fig:convergence-to-liquid}}

  \end{center}
\end{figure}

Fig.~\ref{fig:convergence-to-liquid} shows data for $B=B_\lambda$ and
a higher temperature $t=1/50=0.02$ which is above the melting
temperature. Here, the pancake distribution broadens and consequently
the substrate potential flattens quickly within the first few
substrate iterations. Eventually, the system has become a disordered
liquid with an approximately constant pancake density and the
substrate is virtually flat, as shown for iteration 10. We conclude
that for this temperature and magnetic field the pancake lattice is
unstable to melting into a pancake liquid.

For Fig.~\ref{fig:convergence-to-solid} and
\ref{fig:convergence-to-liquid} we have used 200,000 time-steps for
each substrate-iteration in order to be able to compare the full and
the Fourier-filtered method, but it would be sufficient to use much
less time-steps per substrate-iteration for the Fourier-filtered
methods. For production purposes, we use the \rq Fourier-filtered
method and update the substrate every 500 time-steps
(Sec.~\ref{sec:fourier-filtered-method}).  Although more
substrate-iterations than with the full method are required before the
system reaches a steady state, the \rq Fourier-filtered approach is
more efficient.  The full method and both Fourier-filtered methods
find that at $B=B_\lambda$ the pancake lattice becomes unstable for
$0.017 \leq t \leq 0.018$.

\begin{figure}
  \begin{center}        \includegraphics[clip,width=5cm]{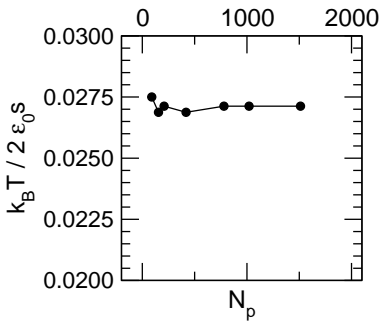}
    \caption{
      Finite-size investigation of instability temperature at
      $B=0.4 B_\lambda$.
      \label{fig:finite-size-plot}}
\end{center}
\end{figure}

Fig.~\ref{fig:finite-size-plot} shows how the instability temperature
varies as a function of system size.  For small numbers of vortices,
$N_\mathrm{p}$, the temperature oscillates slightly and for larger
systems it becomes constant. Most importantly, there is no general
trend visible although the data ranges from $N_\mathrm{p}=90$ to
$N_\mathrm{p}=1512$.  This insensitivity to the system size
demonstrates the local nature of the melting transition at this field.

\subsection{Hysteresis loop}
\label{sec:hysteresis-loop}

\begin{figure}
  \begin{center} \includegraphics[clip,width=8.5cm]{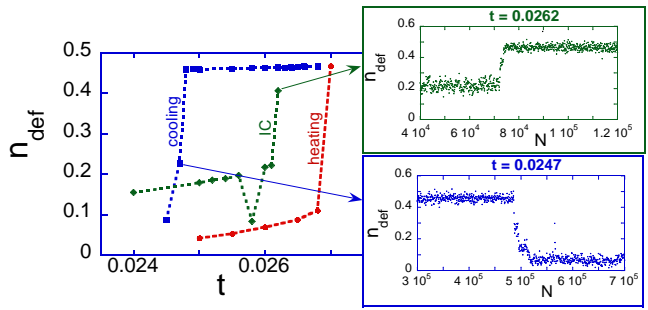}
  \caption{Example of a hysteresis loop obtained by heating  a
    crystal (circles) and cooling a liquid (squares) for $B=0.4
    B_\lambda$. Each point was equilibrated for $9\cdot 10^5$
    Monte-Carlo steps. The crystal melts at $t=0.027$, while the
    liquid freezes at $t=0.0247$. Rhombs represent results obtained by
    simulations starting from the same intermediate defective
    configuration with $n_\mathrm{def}\approx 0.2$, this configuration
    melts at $t=0.0262$, which we take as an estimate for the
    thermodynamic melting temperature. The insets show dependencies of
    the defect concentration on the Monte Carlo step at the
    temperatures where the intermediate configuration melts and the
    liquid configuration freezes.  See also [\onlinecite{movies}].  }
  \label{fig:hysteresis}
  \end{center}
\end{figure}

Rather than starting from a hexagonal crystal for every temperature, a
better approach to determine the instability temperature is to
subsequently increase $t$ until the system melts.  We also find that
by starting from a liquid configuration and lowering $t$, the system
jumps into the crystal state at a certain freezing temperature.  We
expect the true melting temperature to lie within the instability and
the freezing temperature. Such a hysteretic run is shown in
Fig.~\ref{fig:hysteresis}.

In order to estimate the thermodynamic melting temperature at which
the free energy of the solid and the liquid phase cross, we proceed as
follows.  Firstly, we store a vortex configuration taken from 2D
melting-transition simulations. We chose a configuration from a
time-step/sweep where the system was previously a solid but just
starts melting, \emph{i.e.} the defect density starts shooting up and
the structure factor peaks start decaying. This vortex configuration
is ``intermediate'' between a solid and a liquid.  Secondly, we start
the computation from this intermediate configuration (IC) for every
temperature.  The results for the IC simulations are shown in
Fig.~\ref{fig:hysteresis} (rhombs).  We use the temperature at which
the IC melts as a best approximation to the melting temperature of the
physical 3D pancake-vortex lattice.

\subsection{Temperature dependence of substrate curvature and pancake fluctuation width}
\label{sec:alpha}
\begin{figure}
  \includegraphics[width=8cm,clip]{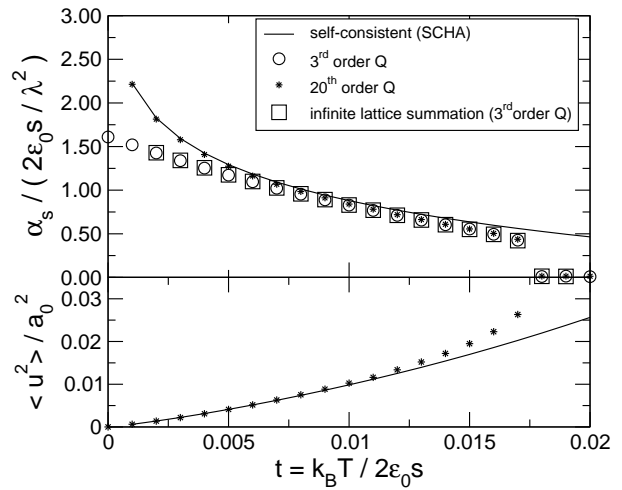}
\caption{
  \emph{Top:} The substrate curvature $\alpha_\mathrm{s}$ at
  $B=B_\lambda$. Shown is our numerical solution of the
  Fourier-filtered method using $Q$-vectors up to 20$^\mathrm{th}$
  order (stars), and using up to 3$^\mathrm{rd}$ order (circles).  We
  have also shown results for $\Qmax = 3 Q_0$ using an infinite
  lattice summation for the in-layer interaction (squares).
  \emph{Bottom:} The pancake fluctuation width $\usqr$ (stars).  For
  comparison, we also show the results of a simple SCHA
  calculation\cite{Dodgson00} (solid line) of the softening of the
  substrate potential (not including thermal softening effects of the
  2D lattice). The results are close to our numerical data at low
  temperatures, but as the melting point is approached there are extra
  (anharmonic) fluctuations in the simulation data for $\usqr$,
  resulting in an even softer substrate potential
  $\alpha_\mathrm{s}$.\label{fig:alpha-and-u2} }
\end{figure}

We can quantify the strength of the substrate potential with
\begin{equation}
  \alpha_\mathrm{s}= \frac{1}{N_\mathrm{p}}\sum\limits_j \frac{\partial^2}{\partial x^2} V_\mathrm{MF}( \mvec{R}_j - \mvec{R}_j^0 ).
\end{equation}
This is the curvature of the potential evaluated at deviations $
\mvec{R}_j - \mvec{R}_j^0 $ from the equilibrium lattice positions $
\mvec{R}_j^0 $ and averaged over pancake positions $ \mvec{R}_j$. The
second derivative can be taken analytically from
(\ref{eq:sparse-method-final-equation-for-substrate}).

Fig.~\ref{fig:alpha-and-u2} shows in the upper plot how
$\alpha_\mathrm{s}$ varies with temperature. The solid line is an
analytical prediction from treating the substrate softening due to
thermal fluctuations within the self-consistent harmonic approximation
(SCHA).\cite{Dodgson00} All other data are simulation results from the
Fourier-filtered method. The stars show $\alpha_\mathrm{s}$ computed
using the \rq Fourier-filtered method with $\mvec{Q}_\mu$-vectors up
to 20$^\mathrm{th}$ order. For low temperatures the data nearly
coincide with the SCHA-solution. Close to melting the SCHA-$\alpha_s$
is larger than the numerical result.  Therefore the simulations give a
softer substrate and the lattice has larger thermal displacements.
This difference could be due to the inadequacies of the SCHA which
does not include the thermal softening of the 2D lattice. The more
complex two-vertex self-consistent harmonic approximation (2VSCHA)
does include these effects.\cite{Dodgson00} The circles in
Fig.~\ref{fig:alpha-and-u2} show results using $\mvec{Q}_\mu$-vectors
up to 3$^\mathrm{rd}$ order, as shown in Fig. \ref{fig:sparse-Qs}.
Close to the transition from solid to liquid around $t\approx 0.0175$
these data agree perfectly with the higher-order data. At lower
temperatures the 3$^\mathrm{rd}$ order results deviate from the
20$^\mathrm{th}$ order because $\usqr$ becomes smaller in the
Debye-Waller factor (\ref{eq:Debye-Waller-factor}). However, as long
as we are interested in temperatures close to the transition, the
3$^\mathrm{rd}$ order approach is sufficient.

The square boxes are computed using the 3$^\mathrm{rd}$ order
approach, but instead of smoothly reducing the pancake
interaction\cite{Fangohr00} at a distance of $\approx 7a_0$, we use an
infinite lattice summation technique for the logarithmic
interaction.\cite{GroenbechJensen96a} This demonstrates that it is
sufficient to use a (smooth) cut-off for the in-layer pancake
interactions.

We compute the average pancake fluctuation width $\usqr$ by fitting to
a distribution where each pancake is normally smeared around its
equilibrium position $\mvec{R}_j^0$
\begin{equation}
  \rho(\mvec{r}) = \frac{1}{2\pi\sigma^2}\sum\limits_j \exp\left(-\frac{|\mvec{r}-\mvec{R_j^0}|^2}{2\sigma^2}\right).
\end{equation}
The Fourier transform of $ \rho(\mvec{r})$ is
\begin{eqnarray}
  \rho(\mvec{q}) &=& \exp\left(\frac{-\sigma^2 q^2}{2}\right) \sum_j  \exp(-i\mvec{q}\sdot\mvec{R}_j^0)\nonumber\\
  &=& (2\pi)^2 n_0 \,\exp\left(\frac{-\sigma^2 q^2}{2}\right) \sum_\mu  \delta^2(\mvec{q}-\mvec{Q}_\mu). \nonumber
\end{eqnarray}
The Fourier components $\rho(\mvec{Q}_\mu)$ have the Debye-Waller
factor as an envelope, and by fitting a Gaussian to it, we can
determine $\usqr=2\sigma^2$.

The lower part of Fig.~\ref{fig:alpha-and-u2} shows computed values
for $\usqr$. We express $\usqr$ in units of $a_0^2$ and it increases
from 0 at zero temperature towards 0.028 close to the transition,
which corresponds to a Lindemann number of $\approx0.168$ at
$B=B_\lambda$.  In agreement with an over-estimation of
$\alpha_\mathrm{s}$ by the SCHA, $\usqr$ is underestimated in
comparison with the numerical results close to the melting transition.

\subsection{Phase diagram}
As demonstrated in Sec.~\ref{sec:convergence}, we can determine for
each parameter pair $(B,T)$ whether the pancake system remains a 3D
pancake lattice, or whether it is unstable towards the liquid phase
which consists of decoupled 2D liquids.  (This is sometimes called a
pancake gas, even though there are still very strong in-plane
correlations in the decoupled layers.  In the absence of Josephson
coupling, a line-like liquid regime is expected only at extremely
small magnetic fields.  \cite{Blatter96})

\begin{figure}
\includegraphics[width=1.0\columnwidth,clip]{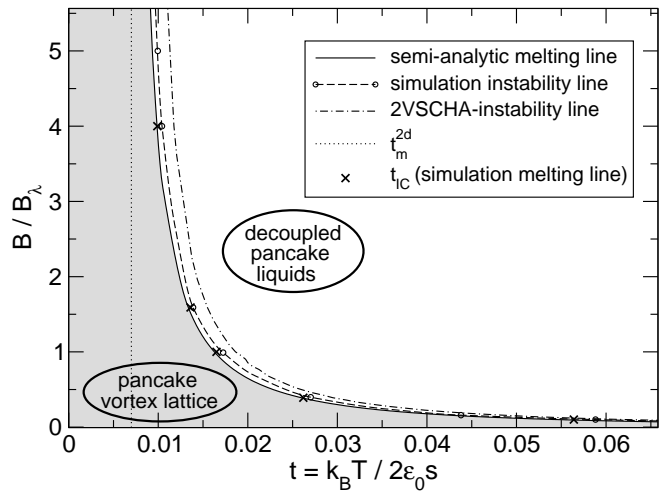}
\caption{
  Phase diagram of the electro-magnetically coupled 3D pancake system.
  Numerically computed instability line (black circles on dashed line)
  in comparison with the instability line from the 2VSCHA (dash-dotted
  line).  Also shown is an semi-analytical estimate for the melting
  line from Ref.\ \onlinecite{Dodgson00}, to be compared with our
  numerical estimate $t_\mathrm{IC}$ (crosses). We have shaded the
  solid phase underneath the melting in gray.  The melting temperature
  $t^\mathrm{2d}_m=0.007$ of a 2D system is shown by a dotted
  line.\label{fig:phasediagram}}
\end{figure}

We probe parameter space in the $B$-$T$-plane as described in Sec.
\ref{sec:convergence} and compute an instability line for the phase
diagram of the system, which is shown in Fig. \ref{fig:phasediagram}
(circles on dashed line).  We also show an estimate of the instability
line that has been computed using the two-vertex self-consistent
harmonic approximation (2VSCHA, dash-dotted) for the substrate
model.\cite{Dodgson00,Dodgson00b} Since in this work we explicitly
compute the pancake positions without using approximations (within the
substrate model), we expect our result to be more accurate than the
2VSCHA. It can be seen that the 2VSCHA slightly overestimates the
temperature for the instability line.

Our numerical estimates for the melting points at certain fields (see
Section~\ref{sec:hysteresis-loop}) are shown as the crosses in
Fig.~\ref{fig:phasediagram}. Also shown as the solid line is the
melting line calculated semi-analytically in
Ref.~\onlinecite{Dodgson00}.  In this work the melting temperature was
estimated by comparing approximate free energies $F=U-TS$ for the
solid and liquid phases.  The solid free energy was calculated from
the SCHA, which gives a variational upper bound on the free energy.
The liquid free energy was taken from earlier simulations of a single
layer,\cite{Caillol82} \emph{i.e. } it was assumed that the layers are
completely uncoupled in the liquid state. Remarkably, our melting
points from simulations lie on top of the semi-analytic line (to
within our error bars).

For increasing fields $B$, the substrate becomes weaker and weaker and
the melting temperature drops. In the limit of $B\to \infty$ we
recover a 2D system with logarithmic interactions for which melting
has been estimated\cite{deLeeuw82,Caillol82,Choquard83} to occur at
$\Gamma_\mathrm{m}^\mathrm{2d} \approx 140\pm 10 \Leftrightarrow
t_\mathrm{m}^\mathrm{2d} \approx 0.007$, which is consistent with our
results.

At low fields the pancake stacks are widely separated and interact
only weakly with each other. In this limit the system melts below the
evaporation transition of an isolated stack of
pancakes\cite{Clem91,Bulaevskii91} at $\Gamma = 4 \Leftrightarrow t =
0.25$. In agreement with this, we find that the instability line
approaches $t\approx 0.25$ for $B\to0$ (see Fig.
\ref{fig:pancakes-in-substrate}).

\subsection{Latent heat and jump in entropy}\label{sec:latentheat}
\begin{figure}
  \begin{center}
      \includegraphics[clip,width=8cm]{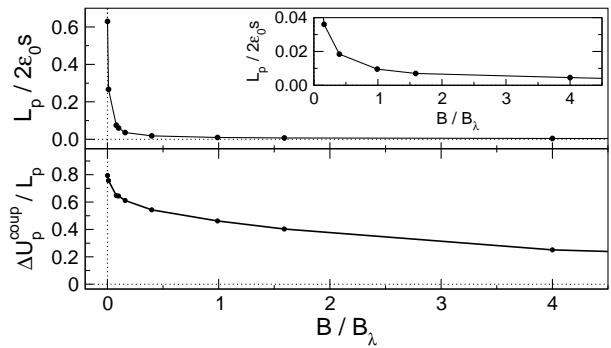}
    \caption{
      \emph{Top:} Latent heat $L_\mathrm{p}$ per pancake across the
      melting transition as a function of field. The inset shows the
      same data on a reduced scale. \emph{Bottom:} Jump in inter-layer
      coupling energy, $\Delta U^\mathrm{coup} = \Ecoup$, normalized
      by latent heat.}
    \label{fig:latent-heat-and-coupling-energy}
  \end{center}
\end{figure}

We compute the latent heat per pancake, $L_\mathrm{p}$, by taking the
difference of internal energy between the solid and the liquid phase
at the melting temperature $T_\mathrm{m}$
\begin{eqnarray}
  \label{eq:latentheat}
  L_\mathrm{p} &=& \frac{1}{N_\mathrm{p}}\left( U_\mathrm{liquid} - U_\mathrm{solid}\right)\\
& =& \frac{1}{N_\mathrm{p}}\left(E^\mathrm{in}_\mathrm{liquid} - ({E}^\mathrm{in}_\mathrm{solid} + \Ecoup)\right).
\end{eqnarray}
The internal energy $U$ of one layer in the solid phase consists of
the in-plane energy $E^\mathrm{in}$
(\ref{eq:in-layer-energy-for-one-layer}) and the inter-layer coupling
energy $\Ecoup$, whereas $\Ecoup=0$ in the liquid phase in our model.
In order to compute $\Ecoup$ for the solid phase, we use
(\ref{eq:E-MF-version2}) where the second sum vanishes due to the
definition of $\delta\rho_n$:
\begin{equation}
  \Ecoup = \frac{1}{2}\int \! \d^2r\: \Vsub(\mathbf{r})
[\rho(\mathbf{r}) - \overline{\rho}], \label{eq:Ecoup-v1}
\end{equation}
where $\overline{\rho}=\Phi_0/B$ is the mean density.
For the Fourier-filtered methods
\begin{eqnarray}
\Ecoup& \stackrel{(\ref{eq:Vsub_FF_1})}{=}& \frac{1}{2}\int \! \d^2r\: \sum\limits_{\mu\neq 0}  {\mathcal U}(\mvec{Q}_\mu) \rho_{\mvec{Q}_\mu} \exp(i\mvec{Q}_\mu\sdot \mvec{r}) \rho(\mathbf{r})   \nonumber \\
 &\stackrel{~(\ref{eq:implementation-rho-q})}{=}& \frac{1}{2} L_xL_y \sum\limits_{\mu\neq 0}  {\mathcal U}(\mvec{Q}_\mu) |\rho_{\mvec{Q}_\mu}|^2.
\end{eqnarray}
For the full method, we have $\rho(\mvec{r})$ as a histogram
available, and we can integrate (\ref{eq:Ecoup-v1}) numerically.

The top plot of Fig.~\ref{fig:latent-heat-and-coupling-energy} shows
how the latent heat varies as a function of field. We have shown the
jump in inter-layer coupling energy normalized by the latent heat in
the bottom part to demonstrate the contribution of the inter-layer
coupling to the latent heat. This plot shows that the substrate
contribution to the latent heat dominates at low fields, and becomes
less and less important towards high fields.

\begin{figure}
  \begin{center}
    \includegraphics[clip,width=7cm]{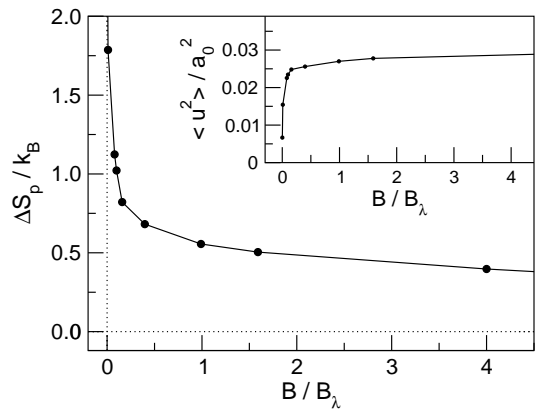}
    \caption{
      Entropy jump $\Delta S_p$ per pancake as a function of field
      $B$.  Inset: Pancake fluctuation width $\usqr$ normalized to
      $a_0^2$ at the melting point (the Lindemann number squared). }
    \label{fig:entropy}
  \end{center}
\end{figure}

Fig.~\ref{fig:entropy} shows the entropy jump across the transition,
$\Delta S_\mathrm{p}\equiv L_\mathrm{p} /T_\mathrm{m}$, as a function
of field.  We find that $\Delta S_\mathrm{p}$ monotonically decreases
with increasing field, as the system approaches the 2D regime. An
important issue is the crossover to the 2D melting regime at very
large $B$. Two melting scenarios are possible in two dimensions: a
usual 1st order melting and continuous dislocation-mediated melting
via the intermediate hexatic phase.\cite{Nelson79,Young79} In the
first case $\Delta S_\mathrm{p}$ has to approach a finite value at
$B\to \infty$ and in the second case it should vanish. Early
simulations for a relatively small number ($\lesssim 500$) of
logarithmically interacting particles suggested a 1st order phase
transition.  \cite{Caillol82,deLeeuw82,Choquard83} However, it is
known that to resolve a continuous melting transition in two
dimensions, very large systems are required (see, e.g., Ref.\ 
\onlinecite{Udink87}).  Therefore, the nature of the melting
transition of 2D particles with logarithmic interactions is an open
issue. Resolving this issue is beyond the scope of this paper.

At low fields the entropy weakly diverges for $B\to0$. We understand
this as follows: the possible configurations scale as $\sim A /
\xi^2$, where $\xi^2$ is the size of a pancake, and $A$ is the space
it can occupy. For the solid state close to the transition the reduced
configuration space is $\sim \usqr / \xi^2$, because the pancake is
confined to an area $A \sim \usqr$. In the liquid the reduced
configuration space grows to $\sim a_0^2 / \xi^2$, where $a_0$ is the
average spacing between pancakes. We get thus an entropy difference
$\Delta S_\mathrm{p}\sim \log( a_0^2 / \usqr )$. Since $\usqr$
approaches at low fields a finite field-independent value of the order
of $\lambda^2$, this explains the observed divergence of $\Delta
S_\mathrm{p}$ at $B\to0$.

For a precise comparison with the experimentally extracted latent heat
of vortex-lattice melting\cite{Zeldov95} one should be careful to
include the temperature dependence of $\lambda$, which was shown in
Ref.\ \onlinecite{Dodgson98a} to give extra terms in the observable
entropy jump.

\section{Conclusions}\label{sec:conclusions}

In this work we have applied a numerical simulation to the suggestion
of Ref.~\onlinecite{Dodgson00} to treat the 3D layered pancake system
with a mean-field approach for the inter-layer interactions, turning
the problem into a 2D system in the presence of a self-consistent
substrate potential. In Ref.~\onlinecite{Dodgson00} this substrate
model was studied with semi-analytic methods.  The 2-vertex
self-consistent harmonic approximation (2VSCHA) was used to estimate
the instability line of the pancake lattice. Also the melting line was
estimated by comparing the elastic free energy of the lattice within
the substrate model to the free energy of independent 2D liquids,
taken from numerical simulations.

Here we have presented results from full numerical simulations of the
substrate model. We have directly calculated both the instability and
melting lines.  Our result for the instability line has a very similar
field-dependence to that of the 2VSCHA, which we find to slightly
overestimate the instability temperature.  The melting line from the
semi-analytic approach agrees within error bars to our simulation
results for the melting of an intermediate configuration. This implies
that the approximations used in Ref.~\onlinecite{Dodgson00} for the
free energies of solid and liquid must be extremely good. We have
computed the pancake fluctuation width $\usqr$ and showed how it
varies as a function of temperature: the variation is significantly
non-linear below the melting transition, as predicted in
Ref.~\onlinecite{Dodgson00}. We also calculate the entropy jump across
the melting transition, which diverges weakly towards small fields and
large melting temperatures.

While we have found a satisfying agreement between our results here
and the earlier approximate work of Ref.~\onlinecite{Dodgson00}, the
true motivation of this project are the possible extensions that can
be studied.  There is now the exciting prospect to study this pancake
vortex system in the presence of pinning disorder. This has been a
controversial topic in recent years
\cite{Khaykovich96,Giamarchi97,Horovitz98,Koshelev98} that our method
should bring some clarity to.

Our results cannot be directly compared with experiments in available
layered superconductors because even in the most anisotropic BSCCO the
Josephson coupling is not negligible. However, the position of the
melting line without the Josephson coupling provides a convenient
reference allowing one to understand the role of the Josephson
coupling in stabilizing the crystalline phase. In particular, it seems
that at low fields even a very small Josephson coupling such as in
BSCCO gives a large upward shift to the melting temperature.  We also
note that it is possible to suppress the effective Josephson coupling
by applying a strong in-plane field, as was done in Ref.\ 
\onlinecite{Mirkovic01}, bringing the melting line closer to the
``Josephson-free'' location.  By extending the model to use several
layers, it is possible to include Josephson coupling between them,
which would realistically describe an anisotropic layered
high-temperature superconductor.

\section{Acknowledgements}
HF thanks Jacek Generowicz for helpful discussions and EPSRC for
financial support.  In Argonne this work was supported by the U.S.
DOE, Office of Science, under contract \# W-31-109-ENG-38.  MJWD is
supported by an EPSRC Advanced Fellowship AF/99/0725.

\appendix

\section{Correlation correction to free energy}
\label{sec:app_free_en_corr}

The correlation correction to the pancake energy, neglected within
the mean-field approach, is given by
\begin{equation}
\delta \mathcal{E}=\frac{1}{2}\sum_{n\neq n^{\prime }}\int \!
\d^2r\,\d^2 r^{\prime} \: U\!\left(
\mathbf{r}-\mathbf{r}^{\prime },n-n^{\prime }\right) \delta \rho
_{n}\left( \mathbf{r}\right) \delta \rho _{n^{\prime }}\left(
\mathbf{r}^{\prime }\right) \label{eq:CorrEnergy}
\end{equation}
The correction to the free energy due to this term
is given up to second order by
\[
\delta \mathcal{F}\approx \left\langle \delta
\mathcal{E}\right\rangle _{0}- \frac{\left\langle \delta
\mathcal{E}^{2}\right\rangle _{0}-\left\langle \delta
\mathcal{E}\right\rangle _{0}^{2}}{2T}
\]
where $\left\langle \ldots \right\rangle _{0}$ implies the
mean-field averaging. Substituting Eq.\ (\ref{eq:CorrEnergy}) in
the last equation and noting that $\left\langle \delta
\mathcal{E}\right\rangle _{0}=0$ we derive for the free energy
correction per pancake, $\delta f\equiv a^{2}\delta \mathcal{
F}/(L_x L_y N)$,
\begin{eqnarray*}
&&\delta f\approx -\frac{a^{2}}{4T L_x L_y}\sum_{n\neq 0}\int\!
\d^2 r\,\d^2 r^{\prime }\!\int\!
\d^2 r_{1}\, \d^2 {r}_{1}^{\prime }\: U\left( \mathbf{r}-
\mathbf{r}^{\prime },n\right)
\\
&&\times U\left( \mathbf{r}_{1}-\mathbf{r}_{1}^{\prime },n\right)
\left\langle \delta \rho \left( \mathbf{r}\right) \delta \rho
\left( \mathbf{r}_{1}\right) \right\rangle _{0}\left\langle \delta
\rho \left( \mathbf{r}^{\prime }\right) \delta \rho \left(
\mathbf{r}_{1}^{\prime }\right) \right\rangle _{0} \end{eqnarray*}
where $N$ is the total number of layers, $L_xL_y$ is the layer area, and $
\left\langle \delta \rho \left( \mathbf{r}\right) \delta \rho
\left( \mathbf{ r}_{1}\right) \right\rangle _{0}\equiv
\left\langle \delta \rho _{n}\left( \mathbf{r}\right) \delta \rho
_{n}\left( \mathbf{r}_{1}\right) \right\rangle _{0}$ is the
density correlation function inside one layer. In the next step we
introduce notation for the sum
\[
W\!\left( \mathbf{r},\mathbf{r}_{1}\right) \equiv \sum_{n\neq
0}U\left( \mathbf{r},n\right) U\left( \mathbf{r}_{1},n\right) ,
\]
for which, using the mixed representation for the interlayer
magnetic interaction, $U(k_{\perp },n)=-{\frac{2\pi
{s}^{2}\varepsilon _{0}}{{\lambda }^{2}{k_{\perp
}^{2}}}}\frac{\exp \left( -ns\sqrt{{\lambda ^{-2}+k_{\perp
}^{2}}}\right) }{\sqrt{{\lambda ^{-2}+k_{\perp }^{2}}}}$, we
obtain the formula
\begin{widetext}
\[
W\!\left( \mathbf{r},\mathbf{r}_{1}\right) =\frac{s^{3}\varepsilon
_{0}^{2}}{2\pi^2\lambda }\int\! \d^2 {k}\,\d^2{k} _{1}\,
{\frac{\exp \left( i\mathbf{k}\sdot\mathbf{r}+i\mathbf{k}_{1}\sdot\mathbf{r}
_{1}\right) }{{k^{2}k_{1}^{2}}\sqrt{{1+\lambda
^{2}k^{2}}}\sqrt{{1+\lambda ^{2}k_{1}^{2}}}\left( \sqrt{{1+\lambda
^{2}k^{2}}}+\sqrt{{1+\lambda ^{2}k_{1}^{2}}}\right) }},
\]
allowing us to represent $W\!\left( \mathbf{r},\mathbf{r}_{1}\right)
$ in a scaling form
\[
W\!\left( \mathbf{r},\mathbf{r}_{1}\right) =\frac{s^{3}\varepsilon
_{0}^{2}}{ \lambda }w\left( \mathbf{r}/\lambda
,\mathbf{r}_{1}/\lambda \right) .
\]
By also using a scaling representation for the in-plane density
correlation function, $\left\langle \delta \rho \left(
\mathbf{r}\right) \delta \rho \left( \mathbf{r}_{1}\right)
\right\rangle _{0}=\frac{1}{a^{4}}h(\mathbf{r}
/a,\mathbf{r}_{1}/a)$, we derive the scaling representation for the
free energy correction
\begin{equation}
\delta f=-s\varepsilon _{0}\frac{s\varepsilon
_{0}}{4T}\frac{s}{\lambda } \mathcal{G}\left[ \frac{a}{\lambda
},\frac{2s\varepsilon _{0}}{T}\right] \label{eq:FreeEnCorrResult}
\end{equation}
where
\begin{equation}
\mathcal{G}\left[ \frac{a}{\lambda },\frac{2s\varepsilon
_{0}}{T}\right] \equiv \frac{1}{L_xL_ya^{6}}\int
\d^2 {r}\,\d^2{r}^{\prime }\,\d^2 {r}_{1}\,\d^2 {r}_{1}^{\prime }
w\!\left( \frac{\mathbf{r}-\mathbf{r}^{\prime }}{
\lambda },\frac{\mathbf{r}_{1}-\mathbf{r}_{1}^{\prime }}{\lambda
}\right) h(
\mathbf{r}/a,\mathbf{r}_{1}/a)h(\mathbf{r}^{\prime }/a,\mathbf{r}
_{1}^{\prime }/a)  \label{eq:RedFunction}
\end{equation}
\end{widetext}
is a dimensionless function of the order unity. The mean-field
free energy per pancake, $f_\mathrm{MF}$, has the scaling property,
\[
f_\mathrm{MF}=s\varepsilon _{0}G\left[ \frac{a}{\lambda
},\frac{2s\varepsilon _{0}}{ T}\right].
\]
Therefore the free energy correction due to interlayer
correlations (\ref {eq:FreeEnCorrResult}) is smaller than the main
term by the factor $ s/\lambda $. In particular, the correlation
correction shifts the melting temperature up as
\begin{equation}
\delta T_\mathrm{m}=\frac{\delta f_{\mathrm{liq}}-\delta
f_{\mathrm{cr}}}{\Delta S} \label{eq:TmShift}
\end{equation}
where $\delta f_{\mathrm{liq}}$ ($\delta f_{\mathrm{cr}})$ is the
correlation correction to the liquid (crystal) free energy at the
melting point and $\Delta S$ is the melting entropy jump. In
principle, the mean-field simulations allow to compute the
correlation correction and the corresponding shift of the melting
temperature. However this computation includes the numerical
evaluation of a quite cumbersome integral in Eq.\ (\ref{eq:RedFunction}).


\begin{thebibliography}{45}
\bibitem[{Fou()}]{FourierDef}
\bibinfo{note}{We define the Fourier transform for a function $f(\mvec{r})$
  depending on the two-dimensional vector $\mvec{r}$ as $ f(\mvec{q}) = \int
  \d^2 r \:\:f(\mvec{r}) \exp( - i \mvec{q}\sdot\mvec{r}) $ and its inverse
  transformation as $ f(\mvec{r}) = (2\pi)^{-2} \int \d^2 q \: \: f(\mvec{q})
  \exp( i \mvec{q}\sdot\mvec{r}) $}.
\end{thebibliography}


\end{document}